\documentclass[12pt,a4paper]{article}
\title{Gauge transformations for a driven quantum particle in 
         an infinite square well}
\author{Stefan Weigert \\
Institut de Physique, Universit\'e de Neuch\^atel\\
Rue A.-L. Breguet 1, CH-2000 Neuch\^atel, Switzerland\\
\texttt{stefan.weigert@iph.unine.ch}}
\date{November 1998}
\newcommand\be{\begin{equation}}
\newcommand\ee{\end{equation}}
\newcommand\bea{\begin{eqnarray}}
\newcommand\eea{\end{eqnarray}}
\newcommand\ket[1]{|#1\rangle}
\newcommand\bra[1]{\langle #1|}

\begin{document}
\maketitle
\begin{abstract}
Quantum mechanics of a particle in an infinite square well under the influence of a time-dependent electric field is reconsidered. In some gauge, the Hamiltonian depends linearly on the momentum operator which is symmetric but not self-adjoint 
when defined on a finite interval. In spite of this symmetric part, the Hamiltonian operator 
is shown to be self-adjoint. This follows from a theorem by Kato and Rellich which guarantees the stability of a self-adjoint operator under certain symmetric perturbations. The result, which has been assumed tacitly by other authors, is important in order to establish the equivalence of different Hamiltonian operators related to each other by quantum gauge transformations. Implications for the quantization procedure of a particle in a box are pointed out.   
\end{abstract}
\subsection*{A. Introduction}
The behaviour of a classical particle in a one-dimensional infinite square-well (a {\em box}, for 
short) under the influence of a time-dependent electric field has been studied in \cite{lin+86}.  The interaction of the charged particle with the field is described by a 
term linear in the position. A time-periodic modulation of this term is sufficient to render the motion of the particle chaotic. 

In quantum mechanics, the system is described by Schr\"odinger's equation on an interval of finite length, with wave functions vanishing at the boundaries. In \cite{eisenberg+97}, it is proposed to apply a gauge transformation to the Hamiltonian operator which results in an interaction term depending linearly on the momentum operator. Since the Hamiltonian operator in this gauge no longer depends on the position, it is straightforward to solve the time-dependent Schr\"odinger equation of this problem. In the classical limit, however, the solutions obtained in this way do not give rise to the expected irregular behaviour. The inconsistency is due to the unjustified assumption that the operator of the kinetic energy commutes with the momentum operator \cite{eisenberg+97}. While they commute for a particle moving on the real axis, they do not commute  if the particle is constrained to move in a box.  

For a particle in a box, a correct quantum mechanical description of the system requires some care in the definition of the operators involved. When acting on wave functions which vanish at the impenetrable walls of the square well, the operator $(\hbar/i) d/dx$ (being the natural candidate for the momentum operator) is {\em symmetric} but {\em not} self-adjoint (cf. Sect. C). {\em A priori}, nothing can be said about its commutator with the self-adjoint operator of the kinetic energy $\propto - d^2/dx^2$. Thus, quantum mechanics on a finite interval provides a simple example where it is necessary to keep in mind that an operator 
is defined ($\imath$) by a prescription of its action on a function {\em and} ($\imath\imath$) by the specification of a domain, i.e., the set of admissible functions.
       
The purpose of this paper is twofold. On the one hand, it aims to clarify the relation between classical and quantum mechanical gauge transformations for a particle living in 
a box (Sect.\ C). On the other hand, the stability of the self-adjointness of the free-particle Hamiltonian in a box under a symmetric perturbation (Sect.\ D) will be shown. This result answers a problem emerging naturally from the discussion in \cite{eisenberg+97} but not been treated there. In the Discussion, the impact of the result on the text-book view on quantization will be discussed. For reference as well as for comparison, gauge transformations for a particle on the real line will be outlined in the following section.        
\subsection*{B. Driven particle on the line}
\label{line}
\subsubsection*{Classical particle}
A classical point particle on the line $ I \!\! R$ subjected to a spatially homogeneous, time-dependent electric field $E(t) = E_0 f(t)$ is described by the Lagrangean 
\be
L_0(x,\dot{x},t) = \frac{m}{2} \dot{x}^2 - \alpha f(t) x \, , \qquad  \alpha = e E_0 \, ,
\label{lagrangeline}
\ee
where $e$ is its charge. In configuration space the particle traces a path  
\be
x(t) = x_{\mbox \tiny hom}(t) + x_{\mbox \tiny in} (t) \, ,
\label{general}
\ee
where $ x_{\mbox \tiny in} (t)$ is any particular solution of the inhomogeneous equation of motion 
\be
\ddot x(t) = - (\alpha/m) f(t) \, ,
\label{eqm}
\ee
while $x_{\mbox \tiny hom}(t)= x(t_0) + (t-t_0) \dot x (t_0)$ is the general solution of the homogeneous equation ($f=0$) associated with (\ref{eqm}). The two  free real parameters are taken as the initial conditions $x(t_0)$ and $\dot{x}(t_0)$ for position and velocity. More explicitly, one can write
\be
x(t) = x(t_0) 
       + (t-t_0) \left( \dot{x}(t_0) + \dot{\xi}(t_0) \right) 
       + \xi(t_0) - \xi(t) \, ,
\label{classsol}
\ee
where the function $\xi(t)$ is given by 
\be
\xi (t) = -\frac{\alpha}{m} \int_{t_0}^t \int_{t_0}^{t'} dt' dt'' f(t'') \, .
\label{doubleint}
\ee
Choosing, for example, a constant function, $f(t) = f_0$, all particles with finite initial 
energy will tend to $-\infty$. Under periodic driving, $f(t) = f_0 \cos(\omega t)$,
most particles will escape to $\pm \infty$, but particles with an initial velocity  
\be
\dot{x}(t_0) = \frac{\alpha f_0}{m \omega}  \sin(\omega t_0)  
\label{specialcondition}
\ee
move back and forth forever in a {\em bounded} region of space since the coefficient multiplying $(t-t_0)$ in (\ref{classsol}) vanishes identically. 

A mathematically equivalent description of this system is given by the Lagrangean  
\be
L_\chi(x,\dot{x},t) 
     = \frac{m}{2} \dot{x}^2 + \alpha \dot{x} F(t) \, ,
       \qquad \frac{d}{dt}F(t) = f(t) \, ,
\label{lagrangetilde}
\ee
which differs from $L_0 (t)$ by a total derivative only: $L_0(t)  =  L_\chi (t) - d\chi(x,t)/dt$. The equation of motion derived from $L_\chi (t)$ is identical to that one following from $L_0(t)$; hence, Eq.\ (\ref{classsol}) still provides the solution. The term $d\chi(x,t)dt$ is physically irrelevant here since the configuration space $ I \!\! R$ of the system is simply connected \cite{morandi92}. Dropping this term corresponds to a gauge transformation which acquires its familiar form in a Hamiltonian description of the system.    

The canonical momenta $p=\partial L / \partial \dot x$ associated with $L_0(t)$ and $L_\chi(t)$, respectively, differ from each other:
\be
p_0 = m \dot{x}  
  \neq m \dot{x} + \alpha F(t) = p_\chi \, .
\label{momenta}
\ee
The corresponding Hamiltonian functions are given by 
\begin{eqnarray}
H_0(x,p_0,t) 
      &=& \frac{p_0^2}{2m}  + \alpha x f(t) \, ,  
\label{hamiltonianzero} \\
H_\chi(x,p_\chi,t) 
      &=& \frac{1}{2m} \left( p_\chi - \alpha F(t) \right)^2 \, ,
\label{hamiltonianchi}
\end{eqnarray}
respectively. The coordinate $x$ is seen to be a cyclic variable in $H_\chi(t)$, hence the momentum $p_\chi$ is a conserved quantity while $p_0$ is not. In Hamiltonian terms, the transition from $H_0(t)$ to $H_\chi (t)$ is effected by applying a gauge transformation  to the electromagnetic potentials: 
\begin{eqnarray}
A_0(x,t) = 0 
     &\to&  A_\chi (x,t) 
            = A_0(x,t) - \nabla \chi (x,t) 
            = - \alpha F(t) \, , \\
\varphi_0 (x,t) = E_0 x f(t) 
     &\to& \varphi_\chi (x,t) 
            = \varphi_0 (x,t) + \frac{\partial \chi(x,t)}{\partial t} 
            = 0 \, , 
\label{gaugefields}
\end{eqnarray}
characterized completely by the function $\chi(x,t) = \alpha x F(t)$. Here one has to keep 
the standard form of the Hamiltonian for a charged particle in mind,
\be 
H(x,p,t) 
      = \frac{1}{2m} \left( p - \frac{e}{c} A(x,t) \right)^2 - e \varphi (x,t) \, .
\label{standard}
\ee
%
%
\subsubsection*{Quantum particle}
Heuristically, quantization of the particle on a line proceeds as follows: position and momentum are replaced by (unbounded) self-adjoint operators $\hat x$ and $\hat p$ which act on of square-integrable complex functions, elements of the Hilbert space ${\cal H} = {\cal L}_2(I \!\! R)$.  The domain of the position operator $\hat x$ is defined by 
\be
{\cal D} ( {\hat x}) 
             = \left\{ \Phi \mid  \Phi \in {\cal L}_2 (I \!\! R) \mbox{ and } 
                               {\hat x} \, \Phi \in {\cal L}_2 (I \!\! R) 
               \right\} \, ,
\label{domainx}
\ee
and ${\cal D} ( {\hat p})$ is defined analogously. These operators satisfy, on an appropriate dense subset \cite{grossmann70} of $\cal H$, the commutation relation  
\be
[ \, \hat p , \hat x ] = \frac{\hbar}{i} \, .
\label{commrel}
\ee
Upon the substitutions $x \to \hat x$ and $p_0 \to \hat p$, the Hamiltonian operator 
associated with $H_0 (t)$ in (\ref{hamiltonianzero}) becomes  
\be
{\widehat H_0} (t) 
      \equiv H_0(\hat x,\hat p,t) 
      =      \frac{{\hat p}^2}{2m}  + \alpha {\hat x} f(t) \, ,
\label{hamiltonianop} 
\ee
acting on states $\ket{ \Phi_0}$ in ${\cal D} ( {\widehat H_0} (t))$ such that $\bra{{\widehat H_0} (t) \Phi_0} {\widehat H_0} (t) \Phi_0 \rangle$ is bounded for all times $t$. Being a linear combination of two self-adjoint operators, the resulting hermitean Hamiltonian can be used as the generator of translations in time using Schr\"odinger's equation:
\be
i \hbar \frac{d}{dt} \ket{\Phi_0 (t)} 
= 
{\widehat H_0} (t) \ket{\Phi_0(t)} \, .
\label{schroeH}
\ee

A second description of the same system is obtained from the quantum counterpart of the Hamiltonian $H_\chi(t)$ in (\ref{hamiltonianchi}) equivalent to $H_0(t)$ up to a gauge transformation. Is determined by applying the {\em same} prescription as before to the second pair of canonically conjugate variables, $x \to \hat x$ and $p_\chi \to \hat p$:
\be
{\widehat H}_\chi (t)
      \equiv H_\chi({\hat x},{\hat p},t) 
      = \frac{1}{2m} \left( {\hat p} - \alpha F(t) \right)^2 \, .
\label{tildehamiltonianop}
\ee
Replacing $\ket{\Phi_0 (t)} \to \ket{\Phi_\chi (t)}$ and ${\widehat H}_0 (t)
\to {\widehat H}_\chi (t)$ in (\ref{schroeH}) yields Schr\"odinger's equation associated with (\ref{tildehamiltonianop}). 

The two descriptions of the quantum particle on the line are equivalent from a physical point of view.\footnote{This is related to the fact that the operators ${\widehat H}_0 (t)$ and ${\widehat H}_\chi (t) $ are both defined unambiguously: no products of non-commuting operators are involved which might give rise to questions of ordering and hence possibly inequivalent Hamiltonians.}  There is a unitary transformation mapping one description to the other. To see this explicitly, consider a quantum mechanical gauge transformation of the state $\ket{\Phi_0(t)}$ effected by the unitary G\"oppert-Mayer operator ${\widehat {\cal U}} (t)$ (\cite{scully+97} App.\ 5.A):  
\be
\ket{{\widetilde \Phi}_0(t)} =  {\cal U}^\dagger (\hat x , t) \ket{\Phi_0(t)} \, , \qquad 
                {\cal U}(\hat x , t) = \exp \left[ - \frac{i}{\hbar} \alpha \hat x F(t)\right] \, . 
\label{unitary}
\ee
Multiply (\ref{schroeH}) with ${\widehat {\cal U}}^\dagger(t)$ from the left which leads to 
\be
i \hbar \frac{d}{dt} \ket{{\widetilde \Phi} (t)} 
  = \left( \frac{1}{2m} \left({\widehat{\cal U}}^\dagger \hat p \, {\widehat {\cal U}}\right)^2  
     + \alpha {\hat x} f(t) - i \hbar {\widehat {\cal U}}^\dagger 
                        \frac{d {\widehat {\cal U}}}{dt}  \right) \ket{{\widetilde \Phi}(t)} 
   =  \frac{1}{2m} \left( {\hat p} - \alpha F(t) \right)^2 \ket{{\widetilde \Phi} (t)} \, ,
%
\label{schroeHtrf}
\ee
where ${\widehat {\cal U}}^\dagger (t) \, \hat p \, {\widehat {\cal U}} (t) = {\hat p} - \alpha F(t) $ has been used. The result is Schr\"odinger's equation associated with ${\widehat H}_\chi$,
hence one can identify: $\ket{{\widetilde \Phi} (t)} \equiv \ket{\Phi_\chi(t)}$. Consequently, the classical gauge transformation (\ref{gaugefields}) has its image in a quantum gauge transformation. In other words, classical and quantum gauge transformations, denoted by $ G_c$ and $G_{qm}$, respectively, and quantization ${\cal Q}$ do commute\footnote{This is not a trivial statement since counterexamples are known \cite{englert98}.} in the present case:
\be
\begin{array}{rcccl}
&&&& \nonumber \\ 
         & H_0            & \stackrel{G_c}{\longrightarrow}    & H_\chi \nonumber  &\\
{\cal Q}: & \downarrow     &                                    & \downarrow        &  \\
         & {\widehat H}_0 & \stackrel{G_{qm}}{\longrightarrow} & {\widehat H}_\chi &.\nonumber\\ 
&&&& \nonumber \\ 
\label{commdiag}
\end{array}
\ee

It is useful for the following to see how this equivalence of the descriptions works out explicitly in terms of the time-evolution operators ${\widehat U}_0(t,t_0)$ and ${\widehat U}_\chi(t,t_0)$. Let us consider a time-independent force, $f(t) = f_0$, for simplicity.  According to Eq. (\ref{schroeH}), the initial state $\ket{\Phi_0(t_0)}$ evolves into 
${\widehat U}_0(t,t_0)\, \ket{\Phi_0(t_0)}$, that is, 
\be
\ket{\Phi_0(t)} = {\cal T} \exp \left[- \frac{i}{\hbar} \int_{t_0}^t dt 
                                       \left( \frac{{\hat p}^2}{2m}  + \alpha f_0 {\hat x}\right)   \right] \ket{\Phi_0(t_0)}
              = \exp \left[- \frac{i}{\hbar} (t-t_0){\widehat H}_0  \right] \ket{\Phi_0(t_0)} \, , 
\label{time}
\ee
where the time ordering denoted by ${\cal T}$ is irrelevant here since ${\widehat H}_0$,
the operator in the exponent, does not depend on time. 

Let us calculate the propagator ${\widehat U}_\chi(t,t_0)$ associated with (\ref{schroeHtrf}). The integrand ${\widehat H}_\chi (t)$ now is explicitly time-dependent. However, time ordering drops out again since the Hamiltonians commute (on a common domain ${\cal D} ({\widehat H}_\chi)$ dense in ${\cal L} ( I \! \! R))$ at different times:
\be
\left[ {\widehat H}_\chi(t), {\widehat H}_\chi(t') \right] 
   = \frac{ \alpha f_0}{m} (t'-t) \left[ {\hat p} , {\hat p}^2 \right] 
   = 0 \, .
\label{Hcommute}
\ee
It is essential here that the operator ${\hat p}^2$ is a function of the operator ${\hat p}$. Using the spectral representation of the momentum operator ${\hat p}$ with eigendistributions $\ket{p}, - \infty < p < \infty$, 
\be
{\hat p} =  \int_{-\infty}^{\infty} dp \ket{p} p \bra{p} \, ,
\label{pspectral}
\ee
one has immediately
\be
{\hat p}^2 =  \left(\int_{-\infty}^{\infty} dp \, \ket{p} p \bra{p}\right)^2 
                =  \int_{-\infty}^{\infty} dp \, \ket{p} p^2 \bra{p} \, ,
\label{p2spectral}
\ee
implying (\ref{Hcommute}). Therefore, in analogy to (\ref{time}), the time evolution of the inital state $\ket{\Phi_\chi(t_0)}$ is given by:
\begin{eqnarray}
\ket{\Phi_\chi(t)} 
  &=& {\widehat U}_\chi(t,t_0) \ket{\Phi_\chi(t_0)}
   = {\cal T} \exp \left[- \frac{i}{\hbar} \int_{t_0}^t dt \,  
                                        {\widehat H}_\chi(t)   \right] \ket{\Phi_\chi(t_0)}\\ \nonumber 
  &=& \exp \left[- \frac{i}{\hbar} 
                \left( (t-t_0) \frac{{\hat p}^2}{2m}  
                        - \frac{\alpha}{2m} (t-t_0)^2 {\hat p}
                        + \frac{\alpha^2}{6m}(t-t_0)^3 
                      \right)\right] \ket{\Phi_\chi (t_0)} \, .
\label{timechi}
\end{eqnarray}
In spite of its slightly complicated appearance this expression is equivalent to (\ref{time}): as shown explicitly in Appendix A, the two solutions are mapped to each other by the operator $\widehat {\cal U} (t)$ in (\ref{unitary}). 

To sum up, the expected analogy between the classical and quantum mechanical description of a driven particle on a line has been established. Classical and quantum mechanical gauge transformations parallel each other. As will be shown in the next section, one has to argue in a more subtle way in order to establish the same result for a particle confined to a box. 
%
%
\subsection*{C. Driven particle in a box}
\label{box}
\subsubsection*{Classical particle}
Suppose now that the particle is confined to a box of lenght $\Lambda$ with boundaries located at $0$ and $\Lambda$, say. 
The time evolution of the particle in between the walls is described by a Lagrangean ${\sf L}_0 (t)$ of the form (\ref{lagrangeline}). It
has to be supplemented by a prescription how to continue the dynamics at the boundaries:
\bea
m \ddot{x} + \alpha f(t) &=& 0 \, , \, \quad \qquad \qquad \qquad \qquad x(t) \in (0, \Lambda) \, , 
\label{newton} \\ 
\left( \dot x (t), x(t) \right) &\to&  \left( -\dot x (t), x(t) \right)   \, , 
                                 \qquad \qquad x(t) = 0 \mbox{ or } \Lambda \, .
\label{reflect}
\eea
Note that the kinetic energy of the particle $\sim {\dot x}^2$ does not change its value upon reflection at a wall. The trajectory in phase space, however, becomes discontinuous\footnote{The discontinuity is due to the simplifying assumption of a truly infinite potential step; one might think of the reflecting boundary at $\Lambda$ as a limit of $V_\sigma(x) = (x/\Lambda)^\sigma, \sigma \to \infty$, giving rise to continuous trajectories for all finite values of $\sigma$.} due to 
$ \dot x (t) \to - \dot x (t)$. 

Let us denote the Hamiltonian of this system by ${\sf H}_0(t) = p^2/2m + \alpha x f(t)$,
to be supplemented by the reflection at the walls:
\be 
\left( p(t) , x(t) \right) \to  \left( -p(t), x(t) \right)   \, , 
                                 \qquad \qquad x(t) = 0 \mbox{ or } \Lambda \, ,
\label{reflectham}
\ee
since velocity $\dot x$ and canonical momentum $p$ are proportional to each other in the present gauge. Therefore, the kinetic energy $\sim p^2$ is seen again to vary continuously as a function of time.

In the absence of an external time-dependent field the motion of the particle is integrable both on the line and in the box: the system has one degree of freedom and the energy provides the required constant of motion. Add now a time-dependent perturbation such as $ \alpha f_0 x \cos (\omega t)$. In the first case, more complicated trajectories result (given in (\ref{classsol})) but the general solution can still be obtained. In the second case, however,   the perturbation renders the motion (now in the box) {\em chaotic} \cite{lichtenberg+92}:
the times of reflection at the boundaries depend sensitively on both the inital values of position and momentum. It is this property which leads to a separation of initially close trajectories at an exponential rate \cite{lin+86}. Effectively, the long-time behaviour behaviour of the system becomes unpredictable.  

Let us turn now to the description of the system in terms of the Lagrangean ${\sf L}_\chi (t)$ in (\ref{lagrangetilde}) which is gauge-equivalent to ${\sf L}_0 (t)$. The equation of motion is given again by (\ref{newton}) as long as the particle moves in between the walls. This part of the dynamics must be supplemented by the condition of reflection, Eq.\ (\ref{reflect}). 

The description of the system acquires a different form, however, when using the Hamiltonian ${\sf H}_\chi (t)= (p- \alpha F(t))^2/2m $. Seemingly, the equations of motion simplify in this gauge since the coordinate turns into a cyclic variable,
\be
{\dot p}_\chi (t) = 0 \, , \qquad 
 \dot x (t) = \frac{1}{m} \left( p_\chi (t) - \alpha F(t) \right) \, , \qquad 
      x(t) \in (0, \Lambda) \, .
\label{chihameq}  
\ee
The quantity $p_\chi^2$, however, is {\em not} a constant of motion: contrary to the velocity $\dot x$, the canonical momentum $p_\chi$ does not just change sign when the particle is reflected at a boundary of the box. In the present gauge, the correct transformation reads  
\be 
\left( p_\chi(t) , x(t) \right) 
        \to   \left( -p_\chi(t) + 2\alpha F(t), x(t) \right) \, ,    
                      \qquad    x(t) = 0 \mbox{ or } \Lambda \, ,
\label{reflecthamchi}
\ee
as follows from (\ref{reflect}) and relation (\ref{momenta}). In this way, the gauge-transformed Hamiltonian ${\sf H}_\chi (t)$ manages to generate the same irregular trajectories as does the original one.  
%
%
\subsubsection*{Quantum particle}
The operator for the kinetic energy of a particle in a box of length $\Lambda$ is defined as 
\be
{\widehat {\sf T}} = \frac{-\hbar^2}{2m} \frac{d^2}{dx^2} \, ,   
\label{kinener} 
\ee
with a domain 
\be
{\cal D} ( {\widehat {\sf T}}) 
             = \left\{ \phi \mid   \phi \in {\cal L}_2 (0,\Lambda) \mbox{ and } 
                               {\widehat {\sf T}} \, \phi \in {\cal L}_2 (0,\Lambda) \, ,
              \phi(0) = \phi(\Lambda) = 0 \right\} \, .
\label{domain0}
\ee
Here ${\cal L}_2 (0,\Lambda)$ denotes the Hilbert space ${\cal H}_\Lambda$  of square integrable functions on the interval $[0,\Lambda]$. The Hamiltonian ${\widehat {\sf T}}$ is a self-adjoint operator as is the  Hamiltonian ${\hat p}^2/2m$ for the free particle on the line.  The solution of the eigenvalue problem ${\widehat {\sf T}} \ket{\phi_n} = E_n \ket{\phi_n}$ is well-known:
\be
\phi_n (x) 
   = \bra{x} \phi_n \rangle
   = \sqrt{\frac{2}{\Lambda}} \sin \left( \frac{n \pi x}{\Lambda}\right) \, , \quad
E_n = \frac{\hbar^2 \pi^2}{2m \Lambda^2} n^2 \, , \qquad n = 1, 2, \ldots
\label{eigenstuff}
\ee
The eigenfunctions $\ket{\phi_n}$ provide an orthonormal basis for the states in the Hilbert space of the system,
\be
\bra{\phi_{n'}} \phi_n \rangle = \delta_{n'n} \, , \qquad 
\sum_{n=1}^\infty \ket{\phi_n}\bra{\phi_n} = 1 \, .
\label{orthonorm}
\ee

How about the momentum operator for the particle constrained to a box? As was pointed out earlier, a self-adjoint momentum operator $\hat p$ exists for the particle on the line, at least in a generalized sense. Surprisingly, an important difference between quantum mechanics on the real line and on a finite interval emerges here. Let us have a look at the natural candidate for the momentum operator on the interval. Consider the operator
\be 
{\widehat {\sf P}} = \frac{\hbar}{i} \frac{d}{dx} \, , 
\label{pbox}
\ee
with domain 
\be
{\cal D} ({\widehat {\sf P}}) 
     = \left\{ \psi \mid \psi \in {\cal L}_2 (0,\Lambda) \, , 
                         \frac{d\psi}{dx} \in {\cal L}_2 (0,\Lambda) \, ,  
              \psi(0) = \psi(\Lambda) = 0 \right\} \, ,
\label{domainDpbox}
\ee
the set of square integrable\footnote{This condition could be replaced by requiring that $\psi$ be absolutely continuous.} functions on the interval $[0,\Lambda]$, vanishing at the boundaries, and with square integrable derivative. The operator ${\widehat {\sf P}}$ has the following properties:\footnote{See \cite{grossmann70} or \cite{reed+72} for details.} it is linear, unbounded, densely defined in ${\cal L}_2 (0,\Lambda)$ and closed. Furthermore, it is a {\em symmetric} operator, that is,
\be
\bra{\psi} {\widehat {\sf P}} \, \phi \rangle 
          = \bra{{\widehat {\sf P}} \, \psi} \phi \rangle \qquad 
                    \mbox{for all } \ket{\phi}, \ket{\psi} \in {\cal D}({\widehat {\sf P}} ) \, ,
\label{symmetric}
\ee
since upon partial integration the boundary terms vanish according to the condition in (\ref{domainDpbox}). It is important that  both the states $\ket{\phi}$ and $\ket{\psi}$ in (\ref{symmetric}) are elements of the domain ${\cal D}({\widehat {\sf P}})$ since
the action of the operator ${\widehat {\sf P}} $ on a state $\ket{\phi} \in {\cal D} ({\widehat {\sf T}}) (\subseteq {\cal D} ({\widehat {\sf P}})$ cf.\ below) generally maps it out of the domain ${\cal D} ({\widehat {\sf T}})$. This is  seen immediately from calculating ${\widehat {\sf P}} \ket{\phi_n}$ using (\ref{eigenstuff}) and (\ref{pbox}), for example. The resulting state does not satisfy the boundary conditions spelled out in (\ref{domainDpbox}): 
\be
( {\widehat {\sf P}} \phi_n ) (0) \neq  0 \neq ( {\widehat {\sf P}} \phi_n )(\Lambda) \, .
\label{boundar}
\ee

Being a symmetric operator, ${\widehat {\sf P}}$ does have an adjoint, ${\widehat {\sf P}}^\dagger$. It is given by $(\hbar/i) d/dx$ on 
\be
{\cal D} ({\widehat {\sf P}}^\dagger) 
     = \left\{ \psi \mid \psi \in {\cal L}_2 (0,\Lambda) \, , 
                         \frac{d\psi}{dx} \in {\cal L}_2 (0,\Lambda) \right\} \, ,
\label{domainDpboxad}
\ee
which differs from ${\cal D} ({\widehat {\sf P}})$ since {\em no} boundary conditions are imposed. Consequently, ${\widehat {\sf P}}^\dagger$ is different from ${\widehat {\sf P}}$, 
or, in other words, the operator ${\widehat {\sf P}}$ is not self-adjoint.

In some cases, a symmetric operator ${\widehat S}$ with domain ${\cal D} ({\widehat S})$ can be `extended' to a self-adjoint operator \cite{reed+72,grossmann70}. This is possible if its defect indices $(m_+,m_-)$ are equal. These two real numbers are determined by the number of linearly independent states in ${\cal D} ({\widehat S})$ which are mapped to zero by the operators $({\widehat S}\pm i)$. Since the defect indices of the symmetric operator ${\widehat {\sf P}}$  are given by (1,1) , it can be promoted to a self-adjoint one, in principle \cite{grossmann70}. 

However, no self-adjoint extension of ${\widehat {\sf P}}$ exists which is compatible with the boundary conditions (\ref{domainDpbox}). The extensions of ${\widehat {\sf P}}$ require a relaxation of the boundary conditions such as {\em periodic} ones: $\psi (0) = e^{ia} \psi (\Lambda)$, $a \in [0,2\pi]$. Whenever $\psi (0)\neq 0$, however, they correspond to a physically different situation, namely to a particle moving on a {\em ring} of length $\Lambda$. Consequently, there is no self-adjoint operator which could be used to describe the momentum of a quantum particle in a box with perfectly reflecting walls. 

The fact that the operator ${\widehat {\sf P}}$ on the interval is symmetric at best, has been used in \cite{eisenberg+97} to resolve an apparent inconsistency when comparing the time evolution of a {\em quantum} particle confined to a box on the one hand and its classical counterpart on the other. The reasoning goes as follows. 

Start from the Hamiltonian operator 
in (\ref{hamiltonianop}) restricted to the interval $[0,\Lambda]$. It originates from a system which is classically nonintegrable: therefore, the associated time evolution will exhibit typical features hinting at the nonintegrability bound to emerge in the classical limit. If the underlying classical system were integrable, no ``quantal fingerprints'' of classically chaotic behaviour could be identified. An equivalent description of the system is possible in terms of the Hamiltonian 
given in (\ref{tildehamiltonianop}), being gauge equivalent to the original Hamiltonian. 
In a sloppy notation the Hamiltonian of the driven particle in the box would read ${\widehat {\sf H}}_\chi^? (t) = (\hat p - \alpha  F(t))^2/2m $. This expression suggests (incorrectly) that the operators of the kinetic energy and of the driving (and hence the Hamiltonians at different times $t$ and $t'$) would commute. Under this (incorrect) assumption, the propagator associated with ${\widehat {\sf H}}_\chi^? (t)$ is immediately calculated in analogy to (\ref{Hcommute}): 
\begin{eqnarray}
{\widehat {\sf U}}_\chi^? (t,t_0) 
  &=& {\cal T} \exp \left[- \frac{i}{\hbar} \int_{t_0}^t dt \, 
                                        {\widehat {\sf H}}_\chi^? (t)   \right]  \nonumber \\
  &\stackrel{?}{=}& \exp \left[- \frac{i}{\hbar} 
                \left(  \frac{{\hat p}^2}{2m} (t-t_0)   
                        - \frac{\alpha}{m}  {\hat p} \int_{t_0}^t dt F(t)
                        + \frac{\alpha^2}{2m} \int_{t_0}^t dt F^2(t)
                      \right)\right] \, .
\label{wrongprop}
\end{eqnarray}
This (incorrect) result, however, produces the propagator of a quantum system with 
{\em integrable} classical counterpart. Contrary to the expectation, no fingerprints of classically nonintegrable motion emerge which is not consistent. 

Upon using the notation introduced here (and drooping the term $\propto F^2 (t)$), the candidate for the Hamiltonian ${\widehat {\sf H}}_\chi (t)$ of a particle in a box reads
\be 
{\widehat {\sf H}}_\chi (t) = {\widehat {\sf T}} - \frac{\alpha}{m}  F(t) {\widehat {\sf P}} \, .
\label{selfsym} 
\ee
It is obvious that the operator ${\widehat {\sf T}}$ of the kinetic energy is {\em not} a function of the momentum operator ${\widehat {\sf P}}$ and, hence, the commutator of ${\widehat {\sf T}}$ and the driving term $(\alpha/m)  F(t) {\widehat {\sf P}}$ is not necessarily equal to zero. It must be evaluated explicitly on an appropriate subset of states  
in the Hilbert space ${\cal H}_\Lambda$. The matrix elements (in the basis (\ref{orthonorm}) of the commutator of the operators ${\widehat {\sf T}}$ and ${\widehat {\sf P}}$ has been evaluated in \cite{eisenberg+97}. Reexpressing the result in terms of the notation used here one obtains
\be 
\bra{\phi_{n^\prime}} \, [ \,  {\widehat {\sf P}} , {\widehat {\sf T}} \, ] \, \ket{\phi_n} 
\propto i n^\prime n \, .
\label{comm}
\ee
Unfortunately, the statement\footnote{The resulting nonassociativity of the momentum operator 
of a particle in a box and the operator of its kinetic energy has been pointed out in \cite{alon+94}.}  
(\ref{comm}) is formal only, since the states  $\ket{\phi_n}$ are mapped out of ${\cal L}_2 (0,\Lambda)$ by the commutator. In any case, there is no reason for the Hamiltonians ${\widehat {\sf H}}_\chi(t)$ and  ${\widehat {\sf H}}_\chi(t')$  to commute for arbitrary times $t\neq t'$. It is not justified to determine the propagator ${\widehat {\sf U}}_\chi (t,t_0)$ as was done in (\ref{wrongprop}). 

Subsequently, Eisenberg {\em et al.} define the quantum dynamics of the system by the operator (\ref{selfsym}) and study its properties in detail, taking into account explicitly the (formally) nonzero commutator of ${\widehat {\sf P}}$ and ${\widehat {\sf T}}$ in (\ref{comm}). The resulting propagator is more complicated, and comparison with the classical counterpart does not lead to inconsistencies any more. 
 
However, a serious gap in the argumantation remains to be filled. Why should the Hamiltonian ${\widehat {\sf H}}_\chi (t)$ in (\ref{selfsym}) provide a trustworthy basis for the description of the quantum particle in a box? A Hamiltonian operator is required to be self-adjoint, otherwise it cannot generate 
a unitary time evolution \cite{reed+72}. As it stands, the operator  ${\widehat {\sf H}}_\chi (t)$ appears to be symmetric only, being a linear combination of the the symmetric operator ${\widehat {\sf P}}$ and the self-adjoint operator of the kinetic energy ${\widehat {\sf T}}$. 

In the following section, the tacitly assumed self-adjointness of the Hamiltonian ${\widehat {\sf H}}_\chi(t)$ in (\ref{selfsym}) will be proved using a basic theorem on the stability of a self-adjoint operator under certain symmetric perturbations. It is this result\footnote{In the mathematical literature a theorem is known which covers the case studied here \cite{weidmann76}. The proof presented here uses elementary methods only.} which will justify its use as generator for the time evolution of the particle in a box.      
%
%
\subsection*{D. Symmetric perturbations of the operator ${\widehat {\sf T}}$}
The self-adjointness of a particle constrained to move in a box is shown to be stable under a wide class of symmetric perturbations. In other words, the self-adjoint part of the Hamiltonian is `stronger' than is symmetric part. The  
situation of interest here, the interaction of the particle with a homogeneous electric field described (in a specific gauge) by a term linear in the operator ${\widehat {\sf P}}$, is covered. 

Consider the `generalized dilation\footnote{For $A(x)=\gamma x$, the operator 
${\widehat {\sf D}}_A$ rescales (dilates) position and momentum operators by constant factors
$\exp \pm \gamma$ \cite{stoler70}.} operator'
\be
{\widehat {\sf D}}_A 
     = \frac{\hbar}{2i} \left( A( x) \frac{d}{dx} +  \frac{d}{dx} A(x) \right) 
\, , 
\label{dilationa}
\ee
where the modulus of both the function $A(x)$ and its derivative are bounded,
\be
| A(x) | \leq A_0 < \infty \, , 
    \quad \left| \frac{dA(x)}{dx} \right| \leq A'_0 < \infty \, , 
    \qquad x \in [0,\Lambda] \, .
\label{dilationb}
\ee
In analogy to (\ref{domainDpbox}), the domain of the operator ${\widehat {\sf D}}_A$ is defined by 
\be
{\cal D} ({\widehat {\sf D}}_A) 
     = \left\{ \psi \mid \psi \in {\cal L}_2 (0,\Lambda) \, ,
                         {\widehat {\sf D}}_A \psi \in {\cal L}_2 (0,\Lambda)\, ,  
              \psi(0) = \psi(\Lambda) = 0 \right\} \, .
\label{domainDA}
\ee
The operator ${\widehat {\sf D}}_A $ is linear, unbounded, densely defined in ${\cal L}_2 (0, \Lambda)$, and closed. In addition, it is symmetric since no boundary terms remain upon partial integration:
\begin{eqnarray}
\bra{\psi} {\widehat {\sf D}}_A \phi \rangle 
 &=&  \left[ \frac{\hbar}{i} \psi^*(x) A(x) \phi (x)\right]_0^\Lambda 
     + \int_0^\Lambda dx 
        \left[ \frac{\hbar}{2i} 
         \left( A(x) \frac{d}{dx} +  \frac{d}{dx} A(x) \right) \psi (x)  \right]^*  
                    \phi (x)  \nonumber \\
&=& \bra{{ \widehat {\sf D}}_A \psi} \phi \rangle  
    \qquad \qquad \mbox{ for all } \ket{\psi}, \ket{\phi} \in {\cal D}({\widehat {\sf D}}_A )\, ,
\label{partialint}
\end{eqnarray}
where the star denotes complex conjugation.

Does ${ \widehat {\sf D}}_A $ have self-adjoint extensions? As shown in Appendix B, it may have 
equal defect indices $m_\pm$ for some functions $A(x)$, while they are unequal for other choices. Only in the first case self-adjoint extensions of ${ \widehat {\sf D}}_A $ do exist. They require, however, periodic boundary conditions just as  
the operator ${\widehat {\sf P}}$. Since they are not appropriate to describe 
a particle in a box, the symmetric dilation operator ${\widehat {\sf D}}_A $ defined on states vanishing at the boundaries cannot be extended to a self-adjoint operator. 
   
Nevertheless, the sum of the self-adjoint ${\widehat {\sf T}}$ and the symmetric ${\widehat {\sf D}}_A $,  
\be
{\widehat {\sf H}}_A 
      = {\widehat {\sf T}} + {\widehat {\sf D}}_A\, ,
\label{generalham} 
\ee
defines a self-adjoint operator with domain ${\cal D} ({\widehat {\sf T}})$ as will be shown now. 
The proof is based on the Kato-Rellich theorem (\cite{reed+72} X.12) for two linear operators ${\widehat M}$ and ${\widehat N}$, densely defined on a Hilbert space $\cal H$ and with ${\cal D}( {\widehat M}) \subseteq {\cal D}( {\widehat N})$. Suppose that the operator ${\widehat M}$ is self-adjoint and that ${\widehat N}$ is symmetric. If the inequality 
\be 
|| {\widehat N} \phi ||^2 
         \leq a \, || {\widehat M} \phi ||^2 + b \, || \phi ||^2 \, ,
            \quad \mbox{ for all } \ket{\phi} \in {\cal D}( {\widehat M}) \, ,
\label{boundedness}
\ee
holds for numbers $a$ and $b$ with 
\be
a < 1 \, , \mbox{ and } b < \infty \, ,
\label{KRcondition}
\ee
then the sum ${\widehat M}+ {\widehat N}$ is a self-adjoint operator with domain ${\cal D}( {\widehat M})$. The perturbation ${\widehat N}$ has to be `sufficiently weak' in order not to destroy the self-adjointness of the operator ${\widehat M}$. In slightly more physical terms: for appropriately chosen constants $a$ and $b$, the expectation value of ${\widehat {\sf D}}_A^2$  in any state $\ket{\phi} \in {\cal D}( {\widehat {\sf T}})$ is required to be smaller than the expectation value of ${\widehat {\sf T}}^2$ in this state plus some finite number proportional to its norm: 
\be 
|| {\widehat {\sf D}}_A \phi ||^2 
         \leq a \, || {\widehat {\sf T}} \, \phi ||^2 + b \, || \phi ||^2 \, ,
            \quad \mbox{ for all } \ket{\phi}  \in {\cal D}( {\widehat {\sf T}}) \, .
\label{bounddilat}
\ee
The proof of (\ref{bounddilat})  proceeds in two steps. First, it will be shown that 
the domain ${\cal D} ({\widehat {\sf D}}_A)$ contains all the states $\ket{{\widehat {\sf T}} \phi}$, 
that is, ${\cal D} ({\widehat {\sf T}}) \subseteq {\cal D} ({\widehat {\sf D}}_A)$. Second, the inequality (\ref{bounddilat}) will be derived. 

Both results follow from the inequality  
\be
 || {\widehat {\sf D}}_A \phi ||^2 
\leq \left( 2 A_0 + A_0^\prime \right) 
  \left( m A_0 \bra{\phi} {\widehat {\sf T}} \phi \rangle + \frac{\hbar^2}{4}  A^{\prime}_0 \right)
%
\label{usefulest} 
\ee
which is proven in Appendix C. Since  
\be
0 \leq  \bra{\phi} \left( {\widehat {\sf T}} - \bra{\phi} {\widehat {\sf T}} \phi \rangle \right)^2 \ket{\phi} 
\qquad \Rightarrow \qquad
\bra{\phi} {\widehat {\sf T}} \phi \rangle^2 
   \leq \bra{\phi} {\widehat {\sf T}}^2 \phi \rangle 
    = || {\widehat {\sf T}} \phi ||  \, ,
\label{positive}
\ee
one concludes from 
\be
|| {\widehat {\sf D}}_A \phi ||^2 
\leq \left( 2 A_0 + A_0^\prime \right) \left( m A_0 || {\widehat {\sf T}} \phi || 
+ \frac{\hbar^2}{4}  A^{\prime}_0  \right) \, ,
\label{domainok} 
\ee
that all states $\ket{{\widehat {\sf T}}^2 \phi}$ with finite norm are automatically elements of ${\cal D} ({\widehat {\sf D}}_A)$, hence:  ${\cal D} ({\widehat {\sf T}}) \subseteq {\cal D} ({\widehat {\sf D}}_A)$.\footnote{In fact, ${\cal D} ({\widehat {\sf D}}_A)$ is even larger than ${\cal D}({\widehat {\sf T}})$ since there are states with finite $\bra{\phi} {\widehat {\sf T}}  \phi \rangle^2$ while $|| {\widehat {\sf T}} \phi || $ is unbounded. Consequently, one has ${\cal D}({\widehat {\sf T}}) \subset {\cal D} ({\widehat {\sf D}}_A)$.}  

Eq.\ (\ref{bounddilat}) is derived by expanding a state $\ket{\phi} \in {\cal D} ({\widehat {\sf T}}) \subset {\cal D} ({\widehat {\sf D}}_A)$ in terms of the basis (\ref{eigenstuff}):
\be
\ket{\phi} 
  = \sum_{n=1}^\infty c_n \ket{\phi_n} \, , \qquad  c_n = \bra{\phi_n} \phi \rangle \, ,
\label{expand}
\ee
leading to 
\be
\bra{\phi} {\widehat {\sf T}} \phi \rangle = \sum_{n=1}^\infty E_n |c_n|^2 
\qquad \mbox{ and } \qquad 
|| {\widehat {\sf T}} \phi ||^2 = \sum_{n=1}^\infty E_n^2 |c_n|^2 \, .
\label{explT}
\ee
The second inequality in (\ref{usefulest}) then reads
\be
|| {\widehat {\sf D}}_A \phi ||^2 
\leq  \left( 2 A_0 + A_0^\prime \right) \left( m A_0  \sum_{n=1}^\infty E_n |c_n |^2
                      + \frac{\hbar^2}{4} A_0^{\prime} ||  \phi ||^2 \right) \, ,
\label{DA2}
\ee
where $||  \phi ||^2= 1$. Since $E_n \sim n^2$, for any positive number $a_0 $ there will exist a finite number $b_0  $ such that 
\be
E_n \leq a_0  E_n^2 + b_0   \, .
\label{estim}
\ee
This allows one to estimate the sum in (\ref{DA2}) by $|| {\widehat {\sf T}} \phi ||^2 $ given in (\ref{explT}): 
\bea 
|| {\widehat {\sf D}}_A \phi ||^2 
&\leq& \left( 2 A_0 + A_0^\prime\right) \left( m A_0 a_0 
         \sum_{n=1}^\infty E_n^2 |c_n|^2 
            + \left(m A_0 b_0    
                      + \frac{\hbar^2}{4} A^{\prime}_0 \right) || \phi ||^2 \right) \nonumber\\
&=& a \, || {\widehat {\sf T}} \, \phi ||^2 + b \, || \phi ||^2 
\qquad \qquad \mbox{ for all } \ket{\phi}  \in {\cal D}( {\widehat {\sf T}}) \, , 
\label{result!}
\eea
where 
\be
a = \left( 2A_0 + A_0^\prime\right) m A_0  a_0  \, , \mbox { and } 
b = \left( 2A_0 + A_0^\prime\right) \left( m A_0 b_0    + \frac{\hbar^2}{4} A^{\prime}_0 \right) \, .
\label{constdef}
\ee
Eq.\ (\ref{result!}) is the required estimate which provides a bound on the perturbation term because one can choose 
\be
a_0   < \frac{1}{\left( 2A_0 + A_0^\prime\right) m A_0} \quad \Rightarrow \quad a<1 \, , 
\label{selectA}
\ee
and the Kato-Rellich theorem applies for an appropriate $b_0  $. 

The operator ${\widehat {\sf H}}_A$ in (\ref{generalham}) has the form of  ${\widehat {\sf H}}_\chi(t)$ in (\ref{selfsym})
if the function $A$ does not depend on $x$, implying $ {\widehat {\sf D}}_A \propto {\widehat {\sf P}}$. As long as the function $F(t)$ remains bounded, ${\widehat {\sf H}}_\chi (t)$ represents therefore a family of self-adjoint operators parametrized by $t$. This justifies the calculations for the driven particle in a box performed in \cite{eisenberg+97} on the basis of the Hamiltonian ${\widehat {\sf H}}_\chi(t)$.   
%
%
\subsection*{E. Conlusions}
It has been shown that the operator ${\widehat {\sf H}}_\chi (t)$ in (\ref{generalham}) provides a trustworthy basis for the description of a quantum particle confined to a finite box. The required self-adjointness, tacitly assumed in \cite{eisenberg+97}, has been proved using a stability theorem by Kato and Rellich. The result sheds some light on gauge transformations in the presence of boundary conditions and on the quantization of a particle in a box.

Already in classical mechanics, gauge transformations have to be handled with care in the presence of boundary conditions. The reflection of the 
particle at the boundary reverses the the sign of the velocity. In one gauge, this behaviour translates into reversing the  canonical momentum while another gauge requires the momentum to transform in a more complicated non-intuitive way. In the first case, upon quantizing the Hamiltonian ${\sf H}_0 (t)$, one ends up with an operator ${\widehat {\sf H}}_0 (t) $ which is manifestly self-adjoint. In the latter case, starting from a classically gauge-equivalent Hamiltonian ${\sf H}_\chi (t)$, a Hamiltonian operator ${\widehat {\sf H}}_\chi(t)$ emerges the self-adjointness of which follows from a detailed analysis only. 

It is satisfactory that, at least for the simple system considered here, one now has two ways to look at the self-adjointness of the operator ${\widehat {\sf H}}_\chi(t)$: on the one hand, it can be thought of being obtained from quantizing directly the classical Hamiltonian ${\sf H}_\chi (t)$; on the other hand one obtains it from quantizing the Hamiltonian ${\sf H}_0 (t)$ in a different gauge, and performing subsequently a quantum mechanical gauge transformation. In other words, quantization ${\cal Q}$ and gauge transformations ($ G_c$ and $G_{qm}$) commute as they did for the particle on the real line:
\be
\begin{array}{rcccl}
& & & & \nonumber \\
         & {\sf H}_0            & \stackrel{G_c}{\longrightarrow}    & {\sf H}_\chi \nonumber  &\\
{\cal Q}: & \downarrow     &                                    & \downarrow        &  \\
         & {\widehat {\sf H}}_0 & \stackrel{G_{qm}}{\longrightarrow} & {\widehat {\sf H}}_\chi &.\nonumber \\
& & & & \nonumber
\label{commdiagbox}
\end{array}
\ee
Since ${\widehat {\sf H}}_0 (t)$ is manifestly self-adjoint, the unitary operator ${\widehat {\cal U}}$, which represents the gauge transformation on the quantum level, is expected to define another self-adjoint operator via ${\widehat {\cal U}}_\chi^\dagger \, {\widehat {\sf H}}_0 \, {\widehat {\cal U}}_\chi$. This argument makes it plausible the self-adjointness of the Hamiltonian ${\widehat {\sf H}}_\chi$ does not come as a surprise.   

It is remarkable that functional analysis seemingly tries to tell us something about the physical world since it it not obvious how to talk (in a quantum mechanical sense) about the momentum of the particle. The boundary conditions (\ref{boundar}) are not compatible with a self-adjoint momentum operator. On the one hand, this is reasonable from a physicist's point of view: no real system allows one to create an {\em infinite} potential barrier. Dropping this idealization by introducing arbitrary large but {\em finite} walls, the system has to be defined on the line -- and a momentum operator with the desired properties {\em does} exist. On the other hand, why should  we need an infinitly extended universe in order to be able to talk about momentum \cite{rosenau98}? 
  
Finally, another interesting point concerns the standard textbook formulation of quantization. Usually, one associates self-adjoint operators $\hat x$ and $\hat p$ with classical canonically conjugate variables. The basic operators are assumed to satisfy Heisenberg's commutation relation, and a Hamiltonian operator is obtained from the classical Hamiltonian by appropriate substitution. However, already the first step of this program cannot be carried out for the particle in the box: no self-adjoint momentum operator exists for the boundary conditions (\ref{boundar}); hence the `algorithm' comes to a halt. This makes it evident that one should not think of quantum mechanics as obtained by a quantization procedure from classical mechanics.  A correct quantum mechanical description may be postulated by  writing down a self-adjoint expression for the Hamiltonian operator of the system such as ${\widehat {\sf H}}_\chi (t)$, for example. To do so, no reference to classical mechanics is needed, in principle.   
\subsubsection*{Acknowledgements}
It is a pleasure to thank Ph.\ Rosenau for a very helpful critical reading of the manuscript as well as M.\ Cibils for a comment. Financial support by the Swiss National Science Foundation is gratefully acknowledged.
\subsection*{Appendix A}
It is shown explicitly that one has 
\bea 
\ket{\Phi_0 (t)}
   &=& {\widehat {\cal U}} (t) \, \ket{\Phi_\chi (t)}
    =  {\widehat {\cal U}} (t)\,  {\widehat U}_\chi (t, t_0) \, \ket{\Phi_\chi (t_0)} \, , \nonumber \\
   &=& {\widehat {\cal U}} (t) \, {\widehat U}_\chi (t, t_0) \, {\widehat {\cal U}}^\dagger (t_0)\ket{\Phi_0 (t_0)}
    =  {\widehat {\cal U}} (t)\,  {\widehat U}_\chi (t, t_0) \, \ket{\Phi_0 (t_0)} \, ,
\label{unitarygauge}
\eea
using ${\widehat {\cal U}}^\dagger (t_0) =1 $. The product of unitary operators is 
equal to  
\begin{eqnarray}
 {\cal U}(\hat x , t)\,  {\widehat U}_\chi (t, t_0)  
    &=& \exp \left[ - \frac{i}{\hbar} (t-t_0) \alpha f_0 \hat x \right] \times \nonumber \\                                                              
& & \times \exp \left[ - \frac{i}{\hbar} 
                      \left( (t-t_0) \frac{{\hat p}^2}{2m}  
                        - \frac{\alpha f_0}{2m} (t-t_0)^2 {\hat p}
                        + \frac{\alpha^2 f_0^2 }{6m}(t-t_0)^3  \right) \right] \nonumber \\
&=& \exp \left[ -\frac{i}{\hbar} (t-t_0) 
                        \left( \frac{{\hat p}^2}{2m}  + \alpha f_0 {\hat x} \right) \right] 
\equiv {\widehat U}_0 (t, t_0) \, , 
\label{equival}
\end{eqnarray}
which involves the Baker-Campbell-Hausdorff relation \cite{wilcox67}:
\be
\exp \left[ {\widehat A} \right] \exp \left[ {\widehat B} \right] 
	     = \exp \left[ {\widehat C} \right] \, ,
\label{compose}
\ee
with 
\be
{\widehat C} = \widehat A + \widehat B 
	       + \frac{1}{2} [\widehat A , \widehat B ] 
	       + \frac{1}{12} \left([\widehat A , [ \widehat A , \widehat B]] 
		      + [[ \widehat A , \widehat B], \widehat B ] \right) 
	       + \ldots
\label{defC}
\ee
In the present case, only the first four terms made explicit here are nonzero, the fifth one and the higher order commutators indicated by the dots all vanish. No qualitative difference arises if one allows for an arbitrary time-dependence $f(t)$.

\subsection*{Appendix B}
The defect indices $(m_+,m_-)$ of the operator ${\widehat {\sf D}}_A $ are defined as the dimensions of the kernels of the operators $({\widehat {\sf D}}_A \mp i )$, that is, the dimensions of the spaces orthogonal to the states $({\widehat {\sf D}}_A \mp i ) \ket{\phi}$ with $\ket{\phi}\in {\cal D} ({\widehat {\sf D}}_A)$ \cite{grossmann70}. These spaces are spanned by states $\ket{\psi}\in {\cal D} ({\widehat {\sf D}}_A)$ which satisfy 
\bea
0 &=& \bra{ \psi} ({\widehat {\sf D}}_A \mp i )\phi \rangle 
   =  \bra{ ({\widehat {\sf D}}_A \pm i ) \psi} \phi \rangle \nonumber \\
  &=& i \int_0^\Lambda dx 
        \left[ \left(  A(x) \frac{d}{dx} +  \frac{1}{2} \frac{d A(x)}{dx} \mp 1 \right) 
                       \psi (x)\right]^* \phi (x) \, , 
\label{defects?}
\eea 
since no boundary terms remain on partial integration. The irrelevant numerical factor $\hbar$ has been set equal to one. The states $\ket{\phi}$ are dense in 
${\cal L}_2 (0,\Lambda)$, hence $\ket{\psi}$ must obey
\be 
A(x) \frac{d\psi(x)}{dx} 
         = \left(\pm 1 - \frac{1}{2} \frac{d A(x)}{dx} \right) \psi(x) \, ,
\label{defectpsis}
\ee
with solutions
\be
\psi_\pm (x) \propto  \frac{ 1}{\sqrt{A(x)}} \exp \left[ \pm 
              \int_x \frac{ dx^\prime}{A(x^\prime)} \right] \, .
\label{solforpsi}
\ee
Suppose now that the function $A(x)$ gives rise to solutions $\psi_\pm (x)$ which are square integrable over the interval $[0,\Lambda]$. Then the defect indices of the operator ${\widehat {\sf D}}_A$ are $(1,1)$ since for each sign $(\pm)$ there is exactly one function orthogonal to the states $({\widehat {\sf D}}_A 
\mp i ) \ket{\phi}$. If one chooses $A(x) = 1$, for example, the result agrees with the defect indices of the derivative on the interval, and (\ref{solforpsi}) reproduces correctly the functions $\psi_\pm (x) \propto \exp [\pm x]$. 

Possible self-adjoint extensions of ${\widehat {\sf D}}_A$ can be obtained only by modifying the boundary conditions. It can be shown that all the  self-adjoint extensions of ${\widehat {\sf D}}_A$ require boundary conditions incompatible with (\ref{boundar}). In analogy to the operator ${\widehat {\sf P}}$ in (\ref{pbox}), the required {\em periodic} boundary conditions correspond to a physically different situation, namely a particle moving on a ring.
\subsection*{Appendix C}
In this Appendix it is shown that the inequality (\ref{usefulest}) holds. In a first step, 
the expression for $|| {\widehat {\sf D}}_A \phi ||^2 $ for a state $\ket{\phi} \in {\cal D} ({\widehat {\sf D}}_A)$ is evaluated:
\bea
\bra{{\widehat {\sf D}}_A \phi} {\widehat {\sf D}}_A \phi \rangle 
   &=& \frac{\hbar^2} {4} \int_{0}^\Lambda dx \left( A( x) \frac{d}{dx} +  \frac{d}{dx} A(x) \right) {\overline \phi}(x) \left( A( x) \frac{d}{dx} +  \frac{d}{dx} A(x)  \right) \phi(x)  \nonumber \\ 
&=& \frac{\hbar^2} {4} \int_{0}^\Lambda dx 
    \left( 4 A^2 {\overline \phi}^\prime  \phi^\prime
         + A^{\prime \, 2} {\overline \phi} \phi 
         + 2 A A^\prime  \{ {\overline \phi}^\prime  \phi 
                          + {\overline \phi}  \phi^\prime \} \right)\, ,
\label{explDA}
\eea
where the argument $x$ is suppressed throughout, the prime denotes the derivative with respect to $x$, and ${\overline \phi}(x)$ is the complex conjugate of $\phi(x)$. The last term can be estimated as follows
\bea
& & \! \! \! \frac{\hbar^2} {2} \!  \int_{0}^\Lambda dx \,
      A A^\prime \left( {\overline \phi}^\prime  \phi 
                         + {\overline \phi}  \phi^\prime \right) 
\leq \frac{\hbar^2} {2} \int_{0}^\Lambda dx \,
      | A A^\prime | \, | {\overline \phi}^\prime  \phi 
                         + {\overline \phi}  \phi^\prime  |   \nonumber \\
& &\leq \hbar^2 A_0 A^\prime_0 \int_{0}^\Lambda dx \,
     | {\overline \phi}^\prime | \, | \phi | 
\leq  \frac{\hbar^2} {2} A_0 A^\prime_0 \int_{0}^\Lambda dx \,
     \left( | {\overline \phi}^\prime |^2 +  | \phi |^2 \right) \nonumber \\
& & = \frac{\hbar^2} {2}  A_0 A^\prime_0 \left( \int_{0}^\Lambda dx \, 
         {\overline \phi}^\prime  \phi^\prime   
        +  \int_{0}^\Lambda dx \, {\overline \phi} \phi \right) 
 =  m A_0 A^\prime_0 \bra{\phi} {\widehat {\sf T}} \phi \rangle 
                  + \frac{\hbar^2} {2} A_0 A^\prime_0 \, ,
\label{herewego}
\eea
using the bounds (\ref{dilationb}) on the functions $A(x)$ and $A^\prime (x)$ and the inequality
\be
0 \leq \left( | {\overline \phi}^\prime (x)| - | \phi^\prime (x)| \right)^2 \, .
\label{standardsquare}
\ee
Similar estimates apply to the first two terms in (\ref{explDA}). After collecting all terms 
one finds 
\be
 || {\widehat {\sf D}}_A \phi ||^2 
\leq \left( 2 A_0 + A_0^\prime \right) 
  \left( m A_0 \bra{\phi} {\widehat {\sf T}} \phi \rangle + \frac{\hbar^2}{4}  A^{\prime}_0 \right) \, ,
%
\label{usefulestapp} 
\ee
which is the required inequality.

\end{document}